\documentclass{article}
\usepackage{graphicx}
\usepackage{amsmath,amssymb}
\usepackage{url}
\usepackage{dcolumn}
\usepackage{adjustbox}
\usepackage{multirow}
\usepackage{color}

\begin{document}

\newcommand{\bogota}[0]{Bogot\'{a} (D.C.)}
\newcommand{\pobflut}[0]{\textit{Poblaci\'{o}n Flotante}}

\title{Report on the \pobflut{} of \bogota{}}

\author{Michele Coscia, Frank Neffke and Eduardo Lora\\CID - Harvard University}

\maketitle

\begin{abstract}
In this document we describe the size of the \pobflut{} of \bogota{}. The \pobflut{} is composed by people who live outside \bogota{}, but who rely on the city for performing their job. We estimate the \pobflut{} impact relying on a new data source provided by telecommunications operators in Colombia, which enables us to estimate how many people commute daily from every municipality of Colombia to a specific area of \bogota{}. We estimate that the size of the \pobflut{} could represent a 5.4\% increase of \bogota{}'s population. During weekdays, the commuters tend to visit the city center more.
\end{abstract}

\section{Introduction}
In this document we aim at describing the size of the \pobflut{} of  Bogot\'{a}, Distrito Capital (hereafter \bogota{}). \bogota{} is a city of about 7.4 million inhabitants\footnote{From census data obtained from the city of \bogota{}, consistent with \url{http://www.dane.gov.co/files/investigaciones/poblacion/proyepobla06_20/Municipal_area_1985-2020.xls}, retrieved on August 18th, 2015.}. Moreover, many people in municipalities outside \bogota{} rely on the city for work and use the services offered. Each person commuting but not residing in \bogota{} is considered part of the \pobflut{} of \bogota{}. A ``commuter'' in this document is defined as an individual who is found more often inside \bogota{} than outside, during working hours (i.e. 7AM to 8PM). It is thought that the \pobflut{} have a large impact on the city, but this impact is difficult to estimate with reliable data. Classical census information provides only a limited picture.

In this document, we use data obtained from telecommunications operators in Colombia. The operators shared cellphone metadata with us that enable us to estimate how many people commute daily from every municipality of Colombia to a specific area of \bogota{}. 

We find that the size of the \pobflut{} relative to the population of \bogota{} is limited. From the cellphone data, we  observe 43,000 cellphones regularly commuting to \bogota{} from the nearby municipalities. Using a rough expansion factor we estimate that the number of commuters is around 400,000, and this represents a 5.4\% impact relative to \bogota{}'s population. There is a high inhomogeneity in the usage of the city by the \pobflut{} from weekdays to weekends. During weekends, the commuting trips are more likely to happen outside working hours. On the other hand, during weekdays the trips tend to disproportionately end in the areas of Chapinero and Santafe. The destinations of the \pobflut{} are overall scattered around the entire territory of \bogota{}. However, when controlling for native population, the highest impact of the \pobflut{} is concentrated on the two main mobility axes of the city: East-West from the airport to Santafe; and North-South from Suba and Usaquen to Chapinero.

\section{Data Description}
The document is based on cellphone call records metadata. The telecommunications operators shared the metadata of all calls that originated from a cellphone operating through their networks during a period of six months. The observation period started in December 1st, 2013 and ended in May 31st, 2014. In total, for the entire nation of Colombia we around 2 billions observations (with one observation being a call). The total number of cellphones observed is larger than 40 millions. The metadata include a variety of attributes for each call: in this paper we focus on the following subset.

\textbf{Originator of the call}. The telecommunications operators provided us with the ID of the cellphone originating the call. Note that this ID is encrypted and anonymized for privacy purposes, rendering the re-identification of the individual impossible. The random IDs in the data are consistent, i.e. the same cellphone is always assigned to the same random ID.

\textbf{Phone Tower used}. To initiate a call, the originator's cellphone has to connect to a cellphone tower. Each cellphone tower is uniquely identified by an ID. We are able to cross this ID with a table, connecting the tower ID to the municipality in which the tower is located. A cellphone cannot be very far from the tower to which it is connecting. This enables us to pinpoint the position of the originator at the moment of the call. We detail our methodology in Section \ref{sec:method}.

\begin{figure}
\centering
\includegraphics[width=.475\textwidth]{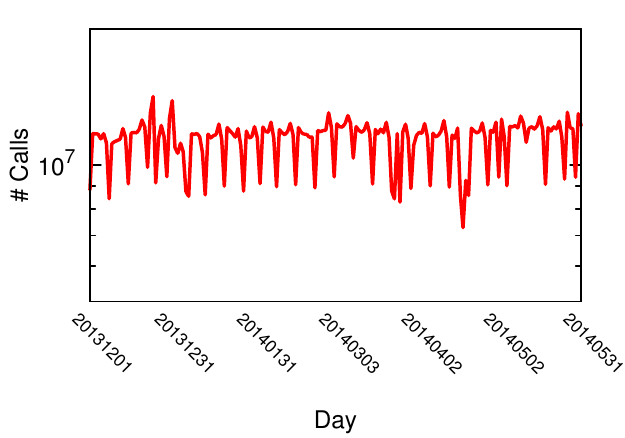}
\includegraphics[width=.475\textwidth]{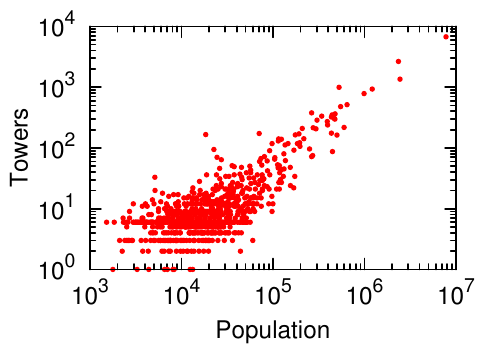}
\caption{\textbf{Statistics on Colombia cellphone usage}. (Left) Number of calls (Y axis) per day (X axis) in our dataset. We cover a time window starting from December 1st, 2013 and ending in May 31st, 2014. The M-shaped pattern is typical of weekly human activities. Deviations are usually represented by national holidays and special occasions, such as New Year's Day. (Right) Number of cellphone antennas against population for each \textit{municipio} in Colombia. \textit{Municipios} with no antennas are removed from the plot.}\label{fig:dailycalls}
\end{figure}

The telecommunications operators which shared the data with us clear on average around 10 million calls every day, with a clear weekly pattern, depicted in Figure \ref{fig:dailycalls} (left).  We can see that the market is pretty stable, without noticeable long term variations. Cellphone coverage in Colombia is inhomogeneous across its municipalities, being directly proportional to the municipality population. Figure \ref{fig:dailycalls} (right) reports the number of cellphone towers per inhabitant. The celltowers increase in a less-than-proportional way: for every 10\% increase in population, there is an 8.5\% increase in celltowers.

\begin{figure}
\centering
\includegraphics[width=.32\columnwidth]{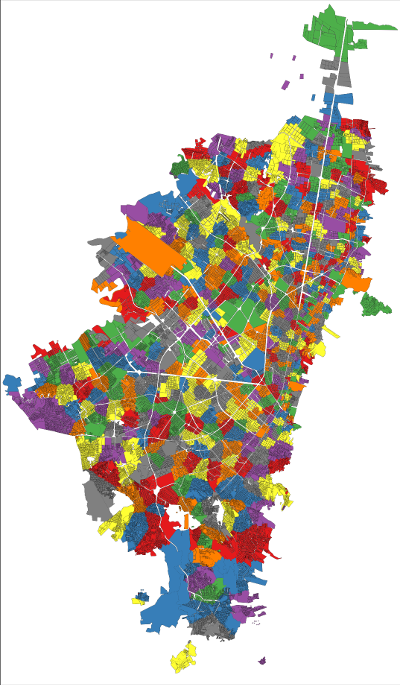}
\includegraphics[width=.32\columnwidth]{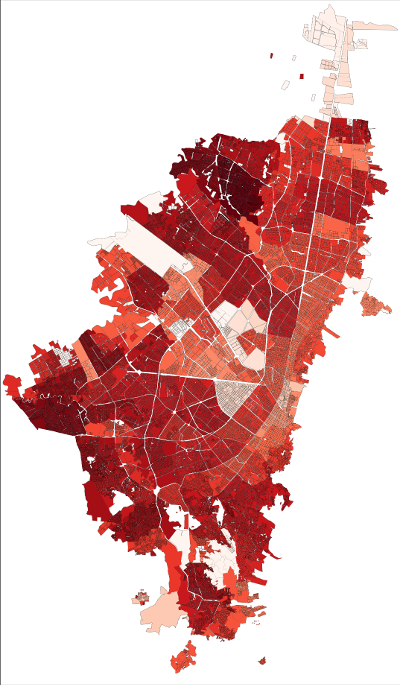}
\includegraphics[width=.32\columnwidth]{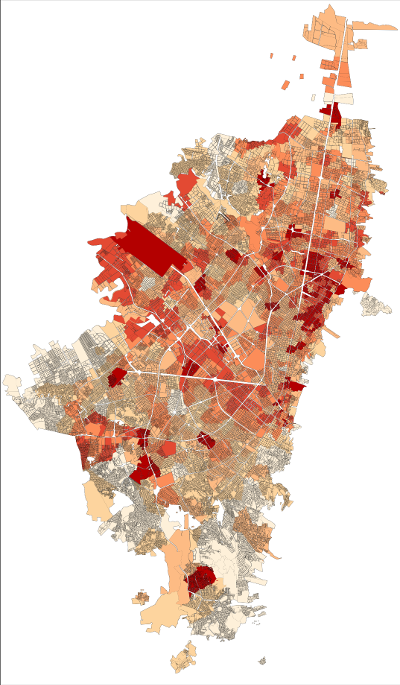}\\
(a)\qquad\qquad\qquad\qquad\qquad(b)\qquad\qquad\qquad\qquad\qquad(c)
\caption{\textbf{Geolocated information on the city of \bogota{}}. (a) We assign each city block (\textit{manzana}) to the cellphone tower coverage area. (b) Population data on \bogota{}, from white (sparsely populated block) to red (densely populated block). Data aggregated at the UPZ level. (c) Productivity of the establishments located in each block. Productivity is calculated as sum of revenues in the block over number of employees, from high productivity (red) to low productivity (white).}
\label{fig:bogota-coverage}
\end{figure}

\bogota{} hosts almost seven thousand phone antennas, and this allows us to subdivide \bogota{} in smaller areas, which will be used later in the document for analysis about the destinations of the \pobflut{}. Multiple antennas can be hosted in the same towers and thus have the same coordinates. The total number of towers in \bogota{} is 650, which is the same number of areas we can define. Each area has been determined by assigning a \bogota{} block to the closest tower, with distances calculated as straight line. Figure \ref{fig:bogota-coverage} (a) reports the resulting \bogota{} areas.

The city of \bogota{} provided us with census and chamber of commerce information about each block of the city. From the census, we obtain information about the population distribution: how many people are registered living in a particular UPZ. A UPZ is a planning unit, that includes several blocks. There are around 100 UPZs in \bogota{}. Figure \ref{fig:bogota-coverage} (b) reports \bogota{}'s population map. From the chamber of commerce we can draw the map of \bogota{}'s productive establishments. In Figure \ref{fig:bogota-coverage} (c), we report the average productivity of establishments in the block. The average productivity is calculated as the total revenue of the establishments in the area divided by the number of people employed.

In the document, we report some statistics about typical commute time for the \pobflut{}. The commute times are calculated using the Google Maps APIs\footnote{\url{https://developers.google.com/maps/documentation/distancematrix/intro}}. The APIs allow us to establish the actual time it takes to drive from two points in space using the actual road path. We estimate the commute time from the center of mass of the origin municipality to the center of mass of \bogota{}. The measure provides the average commute time of the commuters from every origin municipality, rather than evaluating the commute time of each single commuter.

\section{Methodology}\label{sec:method}
The first required step is to assign each observed phone to its home location. This step is necessary because some trips from outside \bogota{} to \bogota{} might have been originated by people living in \bogota{}, unfairly increasing our estimates. This problem has been well studied in the computer science literature and we apply here one of the standard solutions \cite{alexander2015origin}. In practice, we count the number of calls made by each phone connecting to each cell outside working hours. The cell to which the phone connected the most is the most likely to be nearby the home location. In our case, since we are simply interested in the municipality and not in the particular cellphone tower, our estimate is bound to have smaller error.

The telecommunications operators that shared their data with us do not have a full market share in Colombia. Since we can only observe their phones, our calculation will be a severe underestimation. We assume that the sample is random, i.e. the users not covered by the collaborating operators behave, on average, like the average observed user. With this assumption, we can estimate the actual flow from our telecommunications data, defining an expansion factor. The process we used to estimate the expansion factor is detailed in the following section. 

\begin{figure}
\centering
\includegraphics[width=\columnwidth]{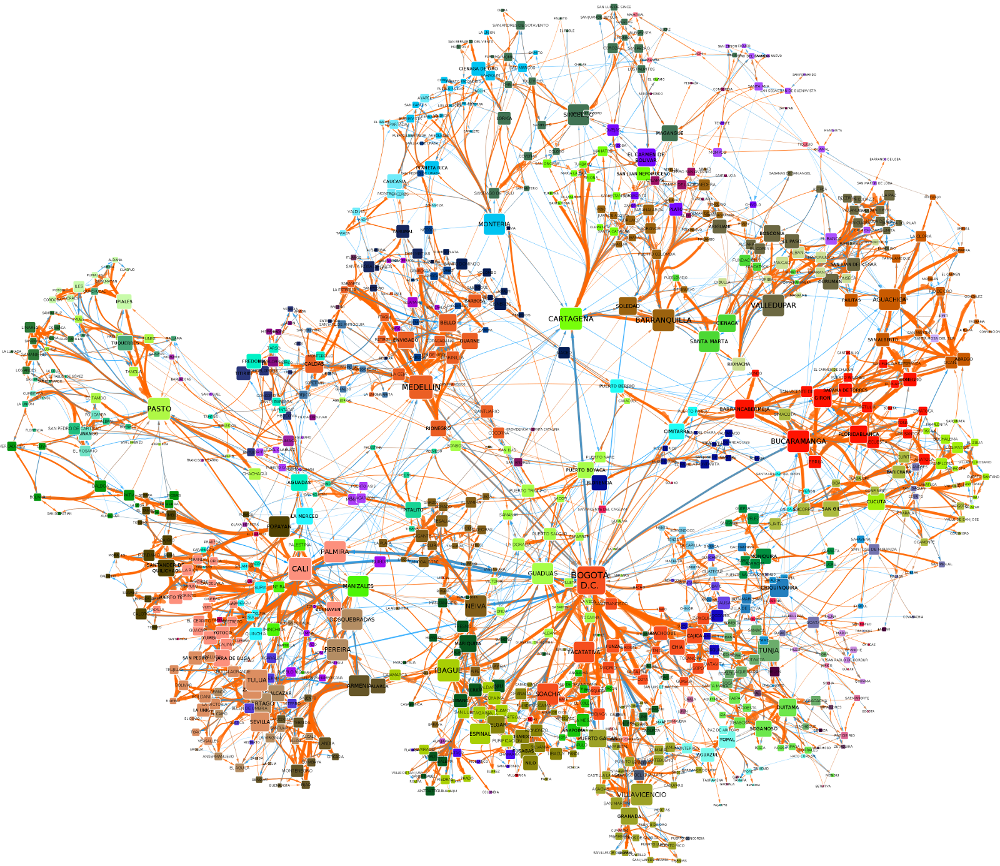}
\caption{\textbf{The mobility network of Colombia}. Each node in the network is a Colombian municipality. Municipalities are connected if a significant number of commuting trips are observed flowing from one municipality to the other. The thickness of the connection is proportional to the number of the commuting trips. The color of the connection is proportional to the significance of the connection, from blue (significant) to red (very significant). Regardless of the color, all connections depicted are significant with $p = 0.01$ or lower. Node size is proportional to the number of municipalities commuting to (in-degree of) the node. Nodes are colored according to their community, i.e. the group of municipalities with which they entertain the strongest connections. Communities are calculated with the Infomap algorithm \cite{rosvall2011multilevel}. We depict the communities on the Colombia map in Figure \ref{fig:signal} (left)}
\label{fig:mobility-network}
\end{figure}

\begin{figure}
\centering
\includegraphics[width=.475\textwidth]{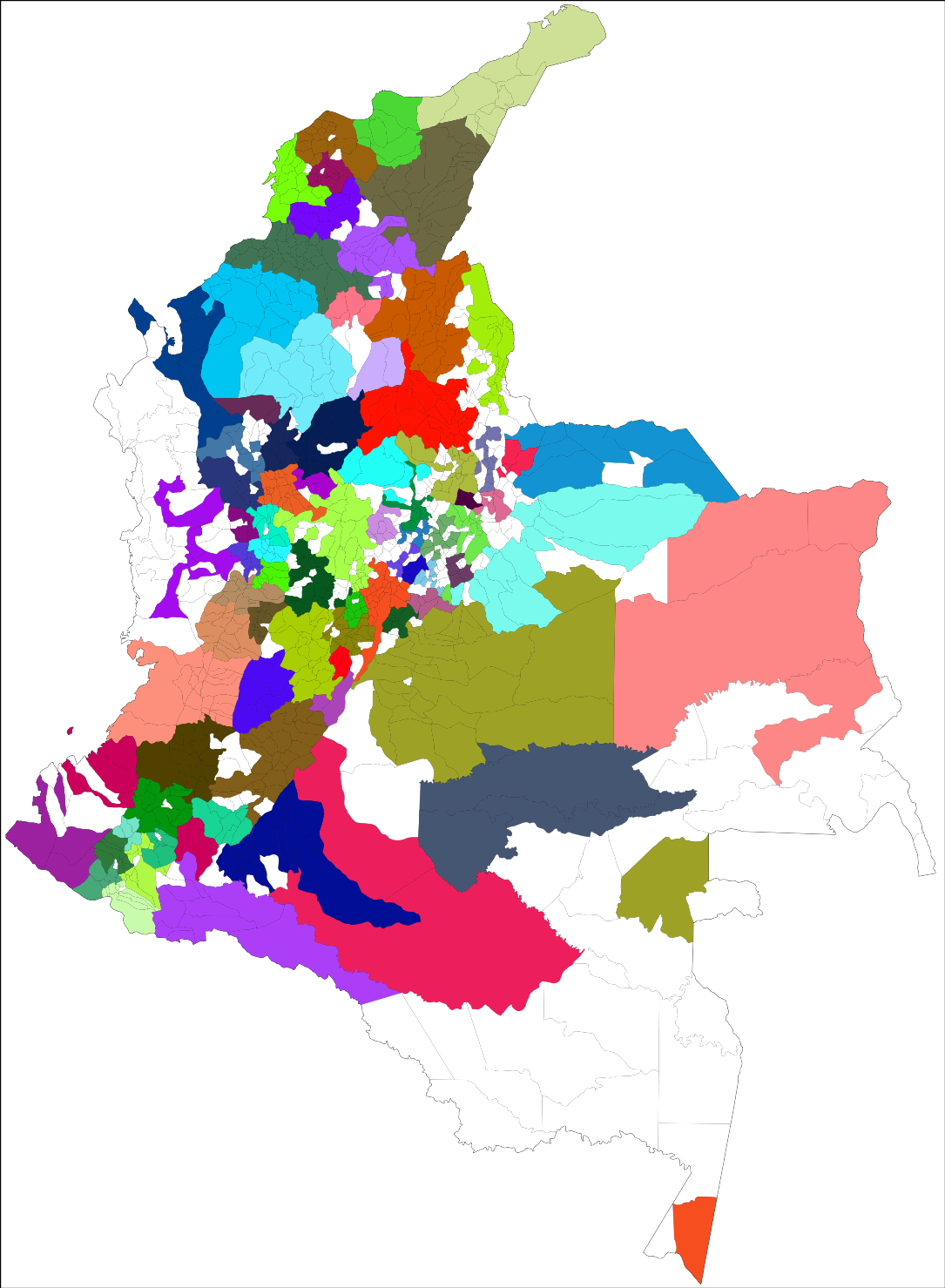}
\includegraphics[width=.475\textwidth]{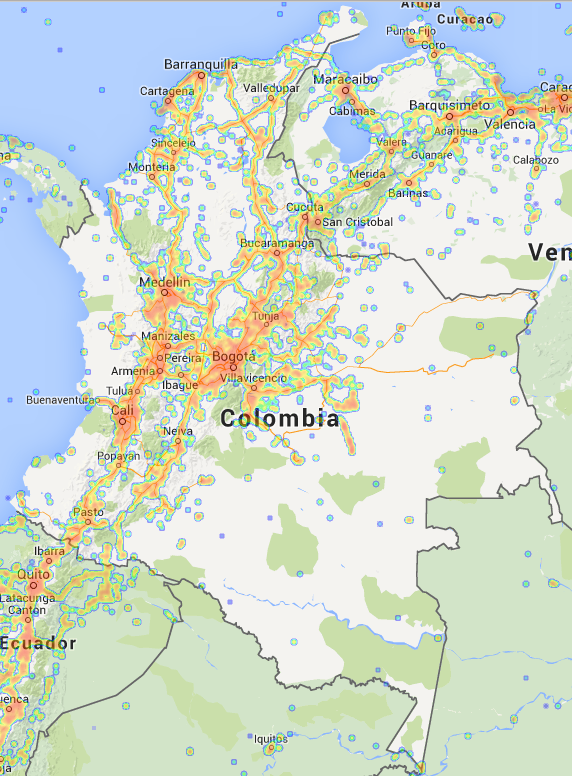}
\caption{\textbf{Colombia territorial coverages}. (Left) A geographical visualization of the network clusters computed on the mobility network. Each municipality area is colored with its corresponding cluster. The color palette is the same used for Figure \ref{fig:mobility-network}. (Right) The heatmap represents the signal strength of the cellphone network across the territory of Colombia. Red areas have a strong signal, blue areas a weak signal, and uncolored map areas have no signal. Image courtesy of \protect\url{opensignal.com}.}\label{fig:signal}
\end{figure}

To generate the mobility network, we keep only IDs which originated and received at least six calls during the observation period. In this way we can drop foreign phones, phones not operating under collaborating operators, and all special phone numbers that are likely to be not associated with an actual person (e.g. call centers). It is important to note that, after this cleaning phase, we do not follow any individual phone number. The networks are aggregated at the level of the municipality. In the case of the mobility network, we connect two municipalities if a caller whose home is in a municipality has been observed in a different municipality. This edge creation criterion is a standard already implemented in several papers creating mobility networks \cite{thiemann2010structure, ratti2010redrawing, coscia2012optimal}. Figure \ref{fig:mobility-network} depicts the resulting mobility network of Colombia. Figure \ref{fig:signal} (left) depicts geographical location of the clusters calculated on the mobility network.


Our last step is to select the municipalities relevant for our analysis. During the six months of observations we have at least one trip to \bogota{} originating from every municipality in Colombia. However, this does not mean that every municipality is part of \bogota{}'s \pobflut{}. To be part of this set, the commute time between the municipality city center and \bogota{} center of mass must be of 180 minutes or less.

One drawback of our methodology is that we cannot describe population coming from a municipality which does not host a cellphone tower. Some municipalities around \bogota{} fall into this category. Figure \ref{fig:signal} (right) reports the cellphone coverage for \bogota{} and we can see that there is no signal in some of the mountain areas around the city. We believe that this issue has a small impact on our study for two reason. First, the uncovered municipalities have small population, thus will contribute little to the \pobflut{}. Second, the cellphone users from those municipalities would still be captured as long as they can connect to a tower in a nearby municipality. We would incorrectly classify their location, but this will help us preserve the accuracy of our estimate in terms of size. 

\section{Expansion \& Validation}


Before being able to use the telecommunications commuting data to analyze the behavior of the \pobflut{} we need to make sure that our commuting estimates are robust. To do so, we focus on the mobility of the inhabitants of \bogota{}, because we have an independent and reliable external dataset for this. We use the data collected for the \textit{Encuesta Movilidad} (EM) of 2011\footnote{\url{http://www.movilidadbogota.gov.co/?pag=954}}. The EM aims at describing how citizens of \bogota{} move around the city: which transportation means they use, from where to where they travel and what is the purpose of their trip.

The data collected by the EM is at the level of \bogota{}'s \textit{Unidades de Planeamiento Zonal} (UPZs), which are a more coarse aggregation than our tower-based division. We aggregate both the UPZs and our subdivision to a higher hierarchical level, the 20 localities (localidades) of \bogota{}. After this aggregation step, we obtain for EM a 20$\times$20 matrix where each cell reports the number of commuters flowing across a specific location pair.

We reconstruct a corresponding matrix also for our observations. We aggregate telecommunication commuting patterns at the UPZ level and we expand this estimate using population data. For each UPZ, say $UPZ1$, we estimate the number of phones living in the area: $c_{UPZ1}$. From EM, we know how many people are resident in the UPZ: $p_{UPZ1}$. The observed trips going from $UPZ1$ to another UPZ ($UPZ2$) are expanded as follows:

$$ T_{UPZ1 \rightarrow UPZ2} = t_{UPZ1 \rightarrow UPZ2} \times \frac{p_{UPZ1}}{c_{UPZ1}},$$

In practice, we calculate an estimated observed market penetration in $UPZ1$ and we use the estimate to weight $t_{UPZ1 \longrightarrow UPZ2}$, which is the raw number of observed phones commuting. We use the same expansion technique for municipalities, substituting municipalities $m_1$ and $m_2$ where appropriate. For municipalities, the population data comes from the National Statistics Department of Colombia (DANE).

\begin{figure}
\centering
\includegraphics[width=.475\textwidth]{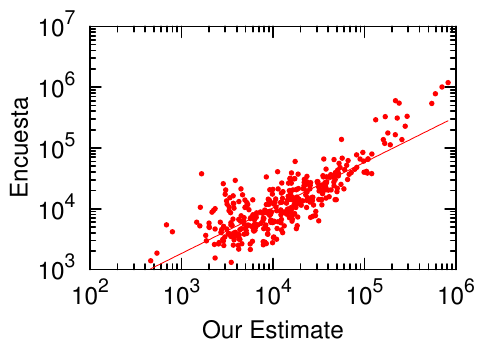}
\caption{\textbf{Cellphone trips and \textit{Encuesta} comparison}. The number of commuting trips according to the elecommunications data analyzed (x axis) against the number of commuting trips estimated by the \textit{Encuesta Movilidad} (y axis) for each pair of \textit{localidades} in \bogota{}.}\label{fig:encuesta-loglog}
\end{figure}

We can now compare the UPZ commuting matrices calculated using EM and telecommunication data. Figure \ref{fig:encuesta-loglog} depicts the relationship between the EM and our estimate. Every observation is a locality pair. In this plot, we expect to find a $\beta$ slope of 1, that is we expect that each mover observation in phone data represents a mover in the EM. However, in our case $\beta \sim 0.75$. A potential explanation for this is measurement error.

\begin{figure}
\centering
\includegraphics[width=.475\textwidth]{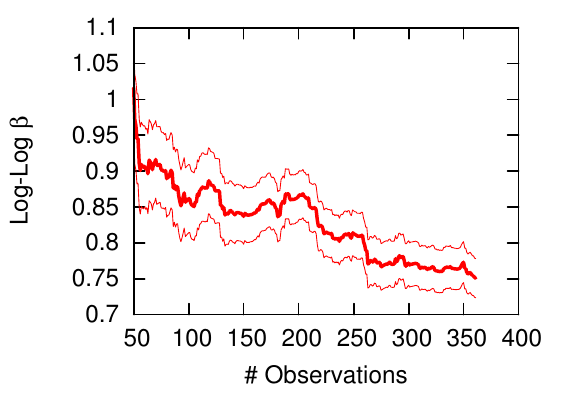}
\includegraphics[width=.475\textwidth]{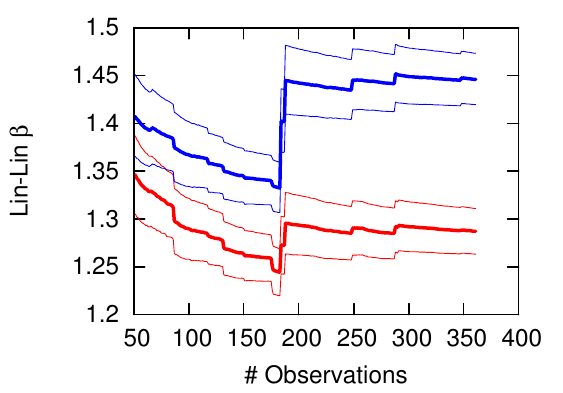}
\caption{\textbf{$\beta$ estimation with progressive noise reduction}. (Left) The evolution of the $\beta$ slope coefficient for the log-log regression between EM and our estimate, with standard error of the estimate. (Right) The evolution of the $\beta$ coefficients for the OLS. The two estimates (blue and red) determine the interval of the actual expansion factor.}\label{fig:betas}
\end{figure}

We progressively remove the observations involving the smallest localities, which would have the most noisy mobility estimations. We sort locality pairs according to their $\sqrt{N_1 N_2}$ value, where $N_x$ is the number of phones residing in locality $x$. We first remove the locality pair with the smallest $\sqrt{N_1 N_2}$ value and we proceed in ascending order. We then recompute the slope of the log-log regression. If the mismatch is due to measurement error, it is likely that there is more error among the smallest places, thus we expect $\beta$ to progressively approach 1 the more noisy connections we remove. Figure \ref{fig:betas}(left) depicts the evolution of the $\beta$ values. Our theory is confirmed, as $\beta$ reaches 1 when we consider only the strongest, and arguably least noisy, connections.

From the log-log regression, we obtain an intercept higher than 1, meaning that our expanded estimate is an underestimation of the actual flows. To confirm this expansion factor, we perform two OLS regressions:

$$ EM = \alpha + \beta TC,$$

$$ TC = \alpha + \beta EM.$$

First we estimate a linear model predicting EM with telecommunication estimate ($TC$), then we take the inverse of the linear model predicting the telecommunication estimate with EM. If the problem is measurement error, the real expansion factor should lie between the two $\beta$ estimates of these models. We again perform the OLS estimations progressively reducing noisy connections. Figure \ref{fig:betas}(left) depicts the interval of our estimation. We can see that the interval narrows down as noise decrease, spanning from 1.35 to 1.4. Hereafter, we decide to use the middle point, and our estimates are expanded by a factor of 1.35.

\section{Totals}
Using the methodology described in the previous section, we are now able to provide the overall estimates of the \pobflut{} size. According to our estimate, on every given day on average fewer than 43,000 observed phones commute to \bogota{}. The 43,000 phones are commuting from 89 Colombian municipalities. In Section \ref{sec:origins} we provide more information about the different origins. Using our expansion factor, we estimate that the \pobflut{} counts more than 400,000 people, and that it increases the population of \bogota{} during the day by $\sim$5.4\%.

\begin{figure}
\centering
\includegraphics[width=.51\textwidth]{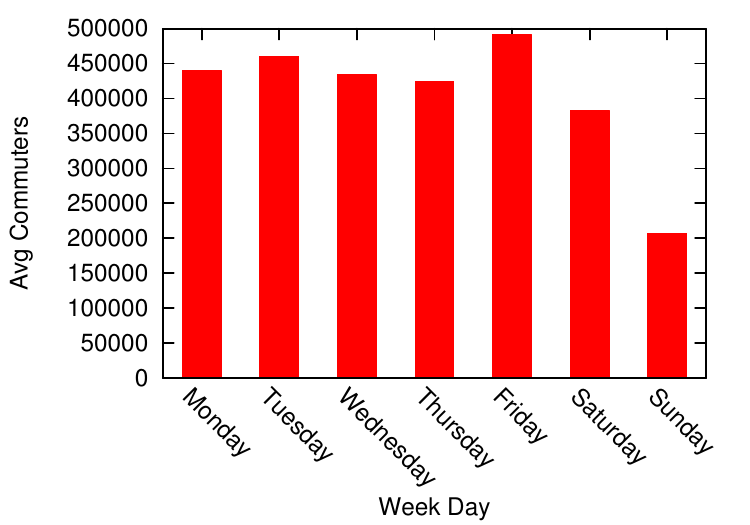}\quad
\includegraphics[width=.45\textwidth]{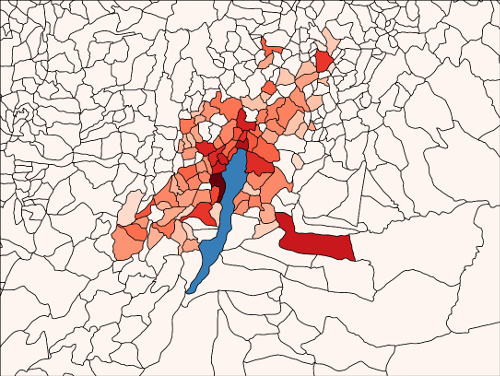}
\caption{\textbf{Commuting relationships with \bogota{}}. (Left) Day of the week averages of our commuter estimates. (Right) Number of commuters from the neighboring municipalities of \bogota{} (highlighted in blue). Darker municipalities contribute most of the commuters. Note that some neighboring municipalities contribute no commuter. That is because those municipalities are not covered by cellphone signal (see Figure \ref{fig:signal}).}\label{fig:commuters}
\end{figure}

The average number of commuters is subject to periodic fluctuations. During the weekend the number of commuters to \bogota{} decreases by one third. Figure \ref{fig:commuters}(left) depicts the weekly pattern in number of commuters during our observation window. During weekdays, the average number of estimated commuters is 450,000, and it goes down to less than 300,000 during weekends.

\begin{figure}
\centering
\includegraphics[width=.5\textwidth]{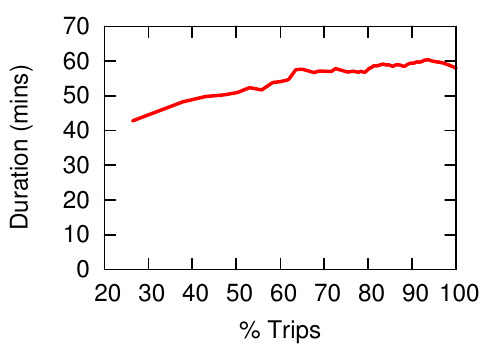}
\caption{\textbf{Trip duration distribution}. The estimated trip duration from the center of the origin municipality to the center of \bogota{}.}\label{fig:duration}
\end{figure}

Using data from Google Maps API we evaluate the duration of typical commute trips from the center of a municipality to the center of \bogota{}. Figure \ref{fig:duration} reports the results. For 50\% of all trips, our estimate of average commute time is 51 minutes. The average commute times touches the hour mark only when we include the farther municipalities. We can conclude that the typical commute time to \bogota{} is less than one hour.

\section{Time Slot Analysis}

\begin{table}
\centering
\begin{tabular}{l|r|rrr}
Day Type & Total & Working Hour & Leisure & Night\\
\hline
Weekend & 0.74 & 1.68 & 0.34 & 0.10\\
Weekday & 1.10 & 1.71 & 0.25 & 0.09\\
\hline
Overall & 1.00 & 1.71 & 0.27 & 0.09\\
\end{tabular}
\caption{\textbf{The normalized share of trips in each time slot}. First we divide all trips according to the day in which they are performed (column ``Total''). Then, out of each subset of trips we report the normalized share of trips happening in each of the three time slots defined in the text.}
\label{tab:timeslots}
\end{table}

We define three different time slots and classify the trips accordingly. The three time slots are: ``Working Hour'', between 7AM and 8PM; ``Leisure'', between 8PM and 12AM; and ``Night'' between 12AM and 7AM. We also keep trace of the day of the week in which each trip is performed: either during a weekday (from Monday to Friday) or during the weekend (Saturday and Sunday). Table \ref{tab:timeslots} reports the normalized share of trips divided in the different time slots. The share is normalized assuming an equal distribution of trips at all hours of the day and for any day of the week. For instance, if trips are equally distributed, we expect to have 28.5\% (2 out of 7) of them in the weekend. Since we observe 21.25\% of trips during the weekend, in Table \ref{tab:timeslots} we report 21.25 / 28.5 = 0.74.

Most of the trips happen during weekdays ($> 1$), suggesting that the primary aim of the \pobflut{} is to work or use the services in \bogota{} rather than visiting the city for pleasure. When breaking down trips in time zones we can confirm this conclusion. A trip made during the leisure or the night time slot is more likely to be performed during the weekend (normalized share of 0.34 vs 0.25).

\begin{figure}
\centering
\includegraphics[width=.32\columnwidth]{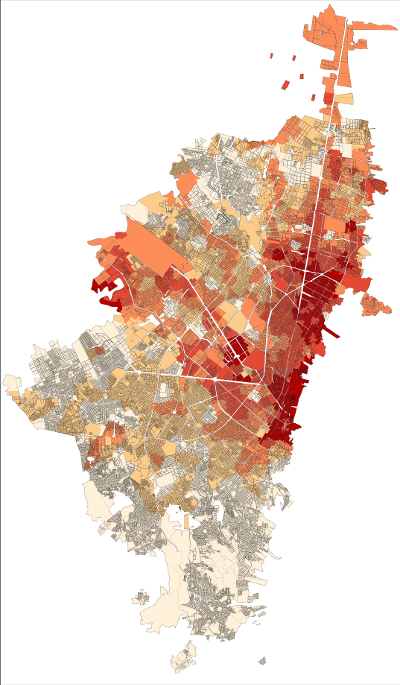}
\caption{\textbf{Weekday vs weekend map}. The difference in destinations from weekdays to weekend. Dark red areas are visited in proportion more during weekdays than during weekends.}
\label{fig:bogota-weekday-difference}
\end{figure}

Figure \ref{fig:bogota-weekday-difference} depicts the difference in destinations inside \bogota{} from weekdays to weekends. The color of each city block is proportional to $\Delta t_d = \log(t_{d}^{we}) - \log(t_{d}^{wd})$, where $t_{d}^{we}$ is the number of trips ending in block $d$ during weekends and $t_{d}^{wd}$ is the number of trips at the same destination $d$ during weekdays. From the figure, we can observe that the locations of Chapinero and Santafe are visited disproportionately more during weekdays, displayed in dark red. During weekends, more trips end in the outskirts of \bogota{} than in these two locations.

\section{Origins}\label{sec:origins}

\begin{table}
\centering
\begin{tabular}{l|rr|r}
\textit{Municipio} & \# Commuters & \% Influx & Median Duration\\
\hline
Soacha & 133,462 & 32.93\% & 43 min.\\
Chia & 37,326 & 9.21\% & 61 min.\\
Villavicencio & 20,448 & 5.05\% & 165 min.\\
Funza & 17,346 & 4.28\% & 62 min.\\
Mosquera & 16,897 & 4.17\% & 62 min.\\
Zipaquira & 15,850 & 3.91\% & 69 min.\\
Facatativa & 13,560 & 3.35\% & 74 min.\\
Madrid & 12,021 & 2.97\% & 63 min.\\
La Calera & 11,539 & 2.85\% & 102 min.\\
Cajica & 10,543 & 2.60\% & 78 min.\\
\hline
Total & 405,293 & 100.00\% & 51 min.\\
\end{tabular}
\caption{\textbf{Top origins}. The ten municipalities contributing the most to the \pobflut{} of \bogota{}. For each municipality we report the average number of observed commuting phones, the share over all commuters and the estimated duration of the trip.}
\label{tab:origins}
\end{table}

In this section we turn our attention to the municipalities from which the \pobflut{} originate. Table \ref{tab:origins} reports the top ten municipalities of origin. Our estimation suggests that a single municipality, Soacha, is responsible for almost a third of the regular commuters to \bogota{}. More than 130,000 commuters on average travel every day from Soacha. Most of the other municipalities present in the list are located in the Cundinamarca department, with the exception of Villavicencio, which is the capital of a department bordering with Cundinamarca (Meta) and thus we expect it to be a larger center gravitating around \bogota{}'s sphere of influence.

We already reported a median travel time of 51 minutes for all commuters. We can see that this is heavily influenced by the connection between Soacha and \bogota{}, whose city centers can be connected in 43 minutes by car. The commute time is higher for the further away municipalities, also in this case led by the capitals of further away departments.

\begin{figure}
\centering
\includegraphics[width=.475\textwidth]{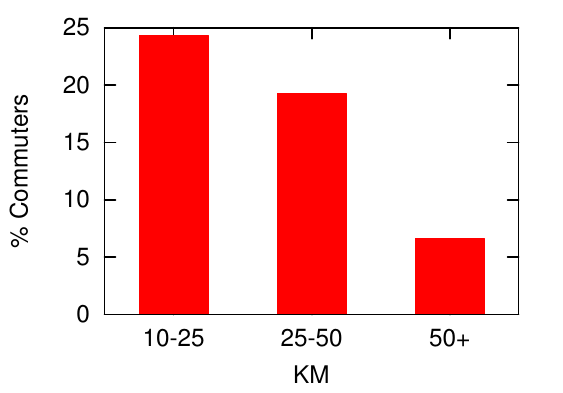}
\caption{\textbf{Share of people commuting a given distance}. For each distance interval, we count the total number of commuters originating from a municipality inside the distance interval and we normalize the count by the combined population of these municipalities.}\label{fig:winkelmann}
\end{figure}

We compare our results with the literature about working commutes. In \cite{winkelmann2010manche} authors study the share of population commuting to work in Germany. They found that around 50\% of the people commute less than 10km, 30\% commute between 10km and 25km, 15\% commute between 25km and 50km and 5\% commute more than 50km. We calculate the corresponding shares in Colombia, using road distance from the municipality of origin to \bogota{}.  

Figure \ref{fig:winkelmann} depicts the result, where we count the number of commuters over the population of the municipality. We cannot calculate the share for the shortest commute, because \bogota{}'s sides are longer than 10km, making us unable to study shorter commutes. However, for the other distance bins we obtain comparable results: almost 25\% commute in the 10-25km interval, less than 20\% commuting in the 25-50km interval, and around 6\% commuting more than 50km.

\section{Destinations}
\begin{figure}
\centering
\includegraphics[width=.32\columnwidth]{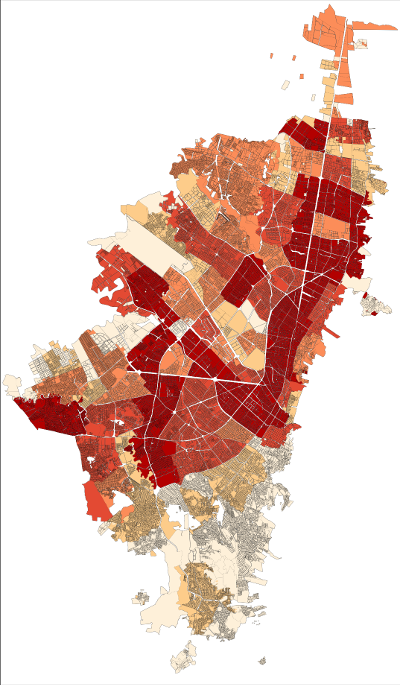}
\caption{\textbf{Destinations of the \pobflut{}}. Each UPZ in \bogota{} is colored according to the average daily influx of commuters from outside the city. Dark red UPZs host a high traffic of commuters, while white UPZs are scarcely used.}
\label{fig:bogota-destinations}
\end{figure}

To estimate the impact of the \pobflut{} on the city it is important to analyze their final destinations. If the commuters target all areas of the city equally their impact is very different than the one they could have if they target specific areas. The first step is to visualize the most popular destinations in \bogota{}. Figure \ref{fig:bogota-destinations} depicts the most popular destinations in red. Figure \ref{fig:bogota-destinations} (right) reports the result. The two main mobility axes of the city host most of the dark red areas: the East-West strip from the airport to Santafe; and the North-South strip from Suba and Usaquen to Chapinero. This means that these axes attract more members of the \pobflut{} compared to the people actually living in the area. In the dark red areas, using our expansion factor, we estimate a commuter over native population ratio of around 0.17, which would imply that the commuters would be a sixth of the registered \bogota{} residents in the area.

\begin{figure}
\centering
\includegraphics[width=.32\columnwidth]{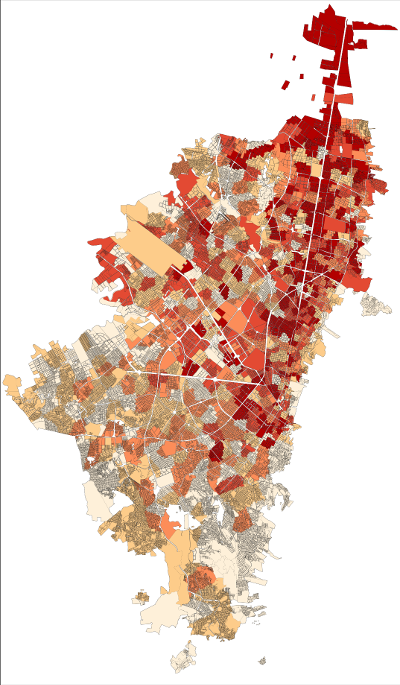}
\includegraphics[width=.32\columnwidth]{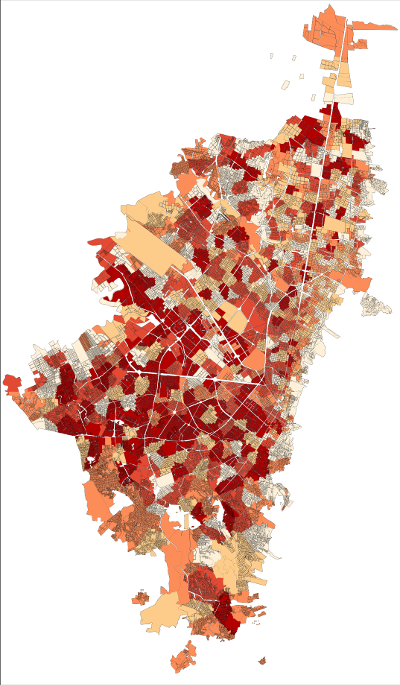}
\caption{\textbf{Different destinations in \bogota{} from commuters in different municipalities}. (Left) Destinations of commuters coming from Chia, from high traffic areas (dark red) to low traffic (white). (Right) Destinations of commuters coming from Fusagasuga, with the same color map.}
\label{fig:bogota-diff-usage}
\end{figure} 

Using our data, we are able to draw maps describing how different municipalities use the city differently. We provide two examples of the many possible. Figure \ref{fig:bogota-diff-usage} reports them. The first one, Figure \ref{fig:bogota-diff-usage}(left), shows the commuter destinations from Chia: popular Chia's destinations are in dark red, while non visited municipalities are in white. We can see that there is a clear range effect. Chia is a municipality bordering \bogota{} on the North, and this is where most of Chia commuters stop.

We can see a similar effect in Figure \ref{fig:bogota-diff-usage}(right), this time showing the destinations of commuters from Fusagasuga, with an identical color map. Fusagasuga is located on the West side. Commuters from Fusagasuga seems to favor the west side of the city, when compared to Chia's destinations.


Using the data from the chamber of commerce, we are able to know the exact city block location of each registered productive establishment in \bogota{}. This allows us to connect the commuting trips to \bogota{} to specific industries. Hereafter, we assume that a commuter will visit an establishment present in the city block at random, proportionally to the establishment's size in number of employees.  


\begin{table}
\centering
\begin{tabular}{l|m{10cm}}
Industry & Description\\
\hline
4330 & Terminacion Y Acabado De Edificios Y Obras De Ingenieria Civil\\
5630 & Expendio De Bebidas Alcoholicas Para El Consumo Dentro Del Establecimiento\\
4711 & Comercio Al Por Menor En Establecimientos No Especializados Con Surtido Compuesto Principalmente Por Alimentos, Bebidas O Tabaco\\
4741 & Comercio Al Por Menor De Computadores, Equipos Perifericos, Programas De Informatica Y Equipos De Telecomunicaciones En Establecimientos Especializados\\
4771 & Comercio Al Por Menor De Prendas De Vestir Y Sus Accesorios (Incluye Articulos De Piel) En Establecimientos Especializados\\
1410 & Confeccion De Prendas De Vestir, Excepto Prendas De Piel\\
4752 & Comercio Al Por Menor De Articulos De Ferreteria, Pinturas Y Productos De Vidrio En Establecimientos Especializados\\
9499 & Actividades De Otras Asociaciones N.C.P.\\
4723 & Comercio Al Por Menor De Carnes (Incluye Aves De Corral), Productos Carnicos, Pescados Y Productos De Mar, En Establecimientos Especializados\\
4111 & Construccion De Edificios Residenciales\\
\end{tabular}
\caption{Top 10 industry of destination during leisure time of weekends that are not in the top 10 in weekdays.}
\label{tab:industry-dest-leis}
\end{table}

Table \ref{tab:industry-dest-leis} reports the top 10 industries of destination during leisure time of weekends. To better highlight the weekdays patterns, we perform several operations. First, we ban from the report the industries that appear in the top 10 according to the previous ranking. Second, we sort each industry using as score the following formula: $WE^2/WD$, where $WE$ is the number of commuters in the weekend and $WD$ is the number of commuters in the weekday. With this score, we discount for the regular weekday activity, while still penalizing the industries that are not visited much neither during weekends nor during weekdays.

As expected, the vast majority of industries in the top 10 is a commercial activity, with bars and pubs ranking in second place.

\section*{Acknowledgements}
We thank the city of \bogota{} for sharing with us the data about population and productive activities in the city, and the shapefiles used for the figures in the document. We thank the telecommunications operators for making the data available to us, and \protect\url{opensignal.com} for allowing us to use one of their maps. We thank Marta C. Gonzales, Serdar Collack, Jameson Toole and Bradley Sturt for their help in gathering, cleaning and hosting the telecommunications data. The Committee on the Use of Human Subjects at the authors' affiliation institution approved the usage of the data made in this paper, certifying that the rights of all subjects whose data have been examined in the study have not been violated.

\bibliographystyle{plain}
\bibliography{biblio}

\begin{thebibliography}{1}

\bibitem{alexander2015origin}
Lauren Alexander, Shan Jiang, Mikel Murga, and Marta~C Gonz{\'a}lez.
\newblock Origin--destination trips by purpose and time of day inferred from
  mobile phone data.
\newblock {\em Transportation Research Part C: Emerging Technologies}, 2015.

\bibitem{coscia2012optimal}
Michele Coscia, Salvatore Rinzivillo, Fosca Giannotti, and Dino Pedreschi.
\newblock Optimal spatial resolution for the analysis of human mobility.
\newblock In {\em Advances in Social Networks Analysis and Mining (ASONAM),
  2012 IEEE/ACM International Conference on}, pages 248--252. IEEE, 2012.

\bibitem{ratti2010redrawing}
Carlo Ratti, Stanislav Sobolevsky, Francesco Calabrese, Clio Andris, Jonathan
  Reades, Mauro Martino, Rob Claxton, and Steven~H Strogatz.
\newblock Redrawing the map of great britain from a network of human
  interactions.
\newblock {\em PloS one}, 5(12):e14248, 2010.

\bibitem{rosvall2011multilevel}
Martin Rosvall and Carl~T Bergstrom.
\newblock Multilevel compression of random walks on networks reveals
  hierarchical organization in large integrated systems.
\newblock {\em PloS one}, 6(4):e18209, 2011.

\bibitem{thiemann2010structure}
Christian Thiemann, Fabian Theis, Daniel Grady, Rafael Brune, and Dirk
  Brockmann.
\newblock The structure of borders in a small world.
\newblock {\em PloS one}, 5(11):e15422, 2010.

\bibitem{winkelmann2010manche}
Ulrike Winkelmann.
\newblock „manche pendeln weit “--berufspendler im
  bundesl{\"a}ndervergleich.
\newblock {\em Statistisches Monatsheft Baden-W{\"u}rttemberg}, 4(2010):40--43,
  2010.

\end{thebibliography}

\end{document}